\documentclass[aps,prd
, twocolumn
,showpacs,superscriptaddress,groupedaddress]{revtex4}
\usepackage{graphicx}
\usepackage{dcolumn}
\usepackage{bm}
\usepackage{amsmath,amssymb}
\usepackage{epsfig}
\usepackage{amsfonts,amsmath,latexsym}
\usepackage[usenames]{color}
\usepackage[usenames,dvipsnames]{xcolor}
\usepackage{type1cm}
\usepackage{eso-pic}
\usepackage{rccol}
\usepackage{pifont}

\def\lsim{\mathrel {\vcenter {\baselineskip 0pt \kern 0pt
    \hbox{$<$} \kern 0pt \hbox{$\sim$} }}}
\def\gsim{\mathrel {\vcenter {\baselineskip 0pt \kern 0pt
    \hbox{$>$} \kern 0pt \hbox{$\sim$} }}}

\begin{document}

\title{Peccei-Quinn violating minimal supergravity and a 126 GeV Higgs}

\author{Csaba Bal\'azs}
\email{Csaba.Balazs@Monash.edu}
\affiliation{%
ARC Centre of Excellence for Particle Physics at the Terascale \\ 
School of Physics, Monash University, Melbourne VIC 3800, Australia.
}
\affiliation{%
Monash Centre for Astrophysics, Monash University, Melbourne VIC 3800, Australia.
}

\author{Sudhir Kumar Gupta}
\email{Sudhir.Gupta@Monash.edu}
\affiliation{%
ARC Centre of Excellence for Particle Physics at the Terascale \\ 
School of Physics, Monash University, Melbourne VIC 3800, Australia.
}

\date{\today}

\begin{abstract}
In a Peccei-Quinn extension of supergravity the $h \to \gamma\gamma$ detection rate can be significantly enhanced due to the reduction of the total Higgs decay width.  To assess the viability of various Peccei-Quinn extensions of minimal supergravity we perform a Bayesian analysis on three such scenarios.  The main constraints on these models come from the currently observed Higgs boson like state by the Large Hadron Collider and from the WMAP observation of dark matter abundance.  Our comparative study reveals that under these constraints the PQ violating scenarios with axino dark matter are clearly preferred over the minimal supegravity model with the lightest neutralino as a dark matter candidate.  
\end{abstract}


\maketitle

\section{\label{pq:intro}Introduction}

The ATLAS~\cite{atlas} and CMS~\cite{cms} collaborations observed an excess in the diphoton invariant mass distribution, around 126 
GeV~\cite{ATLAS-CONF-2012-170,CMS-PAS-HIG-12-045}, with more than $5 \sigma$ statistical significance.  This excess points to a particle with properties close to the 
Standard Model Higgs boson. This evidence is supported by the Tevatron which observed the $b{\bar b}$ decay  mode~\cite{Aaltonen:2012qt,tevathcp}. The 
measured production times decay rates of the particle, for example $ZZ^{(*)}$~\cite{ATLAS-CONF-2012-169,CMS-PAS-HIG-12-041} and $WW^{(*)}$~\cite{ATLAS-CONF-2012-158, 
ATLAS-CONF-2012-162,CMS-PAS-HIG-12-042}, are mostly consistent with those of a standard Higgs. The diphoton rate~\cite{ATLAS-CONF-2012-168, CMS-PAS-HIG-12-015}, however, 
is an exception: it is roughly twice of the standard value.  Although the uncertainties are quite sizable, this anomaly suggest a non-standard Higgs-like particle.  This 
implies that new physics is affecting the properties of the newly discovered particle.

Many new physics models were proposed in the recent literature to explain the difference of the diphoton rate from the standard one.  These models assumed extra 
dimensions~\cite{lhc:extrad}, a fourth fermion generation~\cite{lhc:sm4}, extra vector-like leptons~\cite{lhc:vector}, a dilation~\cite{lhc:dilaton}, extra gauge 
bosons~\cite{lhc:gauge}, or multiple Higgs states~\cite{lhc:multi}.  Solutions based on hybrid models has also been suggested to explain the LHC data in 
Refs.~\cite{lhc:hybrid}.  Data driven studies to estimate the size of the Higgs couplings have also been performed in the Refs.~\cite{hcouth}.  While supersymmetry (SUSY)~\cite{r:susy} is one of the robust new physics candidates, ironically, its minimal versions are struggling to explain the diphoton excess.

In the Minimal Supersymmetric Standard Model (MSSM), for example, it is hard to double the diphoton rate of the lightest Higgs boson.  The most promising scenario relies on the existence of light tau sleptons.  Within the MSSM it is even hard achieve a 126 GeV lightest Higgs boson naturally.  With top squark masses below TeV the lightest MSSM Higgs mass remains below about 120 GeV.  Heavier stop masses, in turn, stretch the hierarchy between the electroweak and SUSY scales.  The problem is severe since every single GeV of loop contribution to the Higgs mass beyond about 120 GeV requires top squark masses further and further above the weak scale.  These problems have been examined in the context of the MSSM~\cite{lhc:mssm} and its simplest extensions, the next-to-minimal MSSM~\cite{lhc:nmssm}, the natural SUSY scenario~\cite{lhc:natural}, and the Peccei-Quinn (PQ) extended MSSM~\cite{blum, lhc:pqnmssm}.  

In a recent work by Blaum {\em et al}~\cite{blum} pointed out that in the models with a slight PQ violation it is possible to raise the Higgs to diphoton decay rates significantly depending on the size of the PQ violating couplings in the Higgs potential.  It is also expected that these models can serve better from the point of view of cosmology by providing an alternative to the traditional neutralino dark matter and leptogensis~\cite{pqcosmo}.  With this motivation, in the current work, we analyse the PQMSSM in greater detail using a Bayesian framework and compare our findings with the MSSM to show that the PQMSSM scenarios fit the collider and cosmological data with more flexibility.  

The organisation of this paper is as follows.  Section \ref{pq:model} provides a brief overview of the PQMSSM and discusses its advantages over the MSSM.  Later in the same section also discusses implications on collider phenomenology and cosmology.  Section \ref{pq:stat} contains our detailed Bayesian analysis of various PQMSSM scenarios.  Our results and conclusions are given in Sections 
\ref{pq:res} and \ref{pq:concl}, respectively.

\section{\label{pq:model} Peccei-Quinn violation in the MSSM}

To solve the strong CP problem Peccei and Quinn (PQ) extended the SM with a global $U(1)$ symmetry \cite{r:pqv1, r:pqv2}.  The $U(1)_{PQ}$ symmetry is spontaneously broken at a scale $\Lambda_{PQ}$ and the pseudo-Goldstone boson induced by this breaking is the axion.  The axion mass is related to the symmetry breaking scale as
\begin{eqnarray}
 m_a \simeq \frac{6.2\times 10^{-3}}{\Lambda_{PQ}} {~\rm GeV} .
\end{eqnarray}
The PQ extension of supersymmetric models requires the addition of a chiral superfield 
\begin{eqnarray}
 \hat \Phi_a = \frac{s + i a}{\sqrt{2}} + \theta \tilde a + \theta \bar\theta F_a , 
\end{eqnarray}
where $s$ is the scalar axion or saxion, $a$ is the the pseudo-scalar axion, $\tilde a$ is the axino, and $F_a$ is an auxiliary field.  The scalar and pseudo-scalar axion fields are even while the axino field is odd under the R-parity.  The masses of these field depend on the supersymmetry breaking mechanism.  In most cases the saxion is ultra heavy with a mass of about $\Lambda_{PQ}$ while the axino mass is highly model dependent.  For the supegravity inspired model, which is subject of the current paper, the axino mass takes the following form\cite{axinomass}, 
\begin{eqnarray}
 m_{\tilde a} \simeq \eta^\kappa M_{Pl} , \\
 {\rm with~} \eta = \frac{\Lambda_{PQ}}{M_{Pl}} {~~\rm and ~~} \kappa \gsim 2 .
\end{eqnarray}
Thus depending on the value of $\kappa$ and $\Lambda_{PQ}$, $m_{\tilde a}$ can take values from a few eV to several TeV.  This wide range of axion masses leads to interesting cosmological consequences which will be discussed in Section \ref{pq:cosmo}.

A broken PQ symmetry contributes to the neutron electric-dipole moment (nEDM) at tree level \cite{r:pqv2, nedm}.  The current experimental limit on the nEDM is 
\begin{eqnarray}
 d_n < {\left| 1.9\times 10^{-26}\right|} ~e\,{\rm cm} , 
\end{eqnarray}
at 90\% CL.  
This limit translates into a lower bound on the PQ breaking scale.  A model dependent upper bound has also been obtained for the PQSuGra case in Ref.~\cite{pqrelic1}, leading to
\begin{eqnarray}
 1 \times 10^9 ~{\rm GeV} < \Lambda_{PQ} < 5\times 10^{11} ~{\rm GeV} .
\label{Eq:LambdaPQBounds}
\end{eqnarray}

\begin{figure}
\centerline{\includegraphics[angle=-90, width=.50\textwidth]{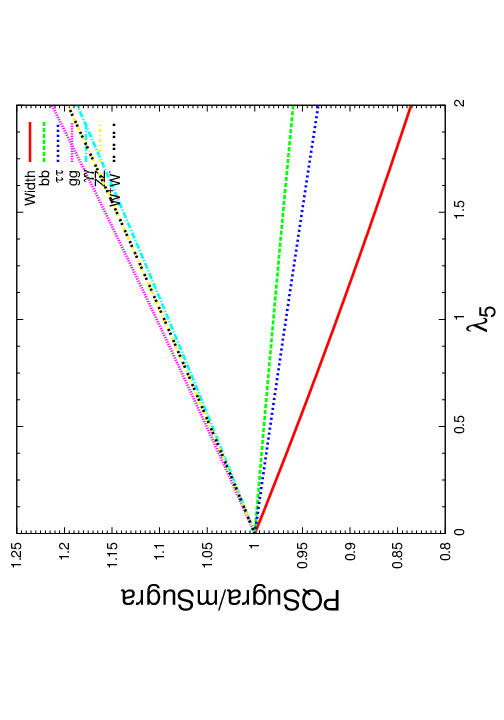}}
\caption{\color{black!100}{\sf Ratio of Total decay width and various Branching ratios of the SM-like Higgs boson in the PQ-violating mSuGra to the conserved mSuGra case for $m_0 = 1000, m_{1/2} = 500, A_0 = 0$ (all in GeV units), $\tan\beta = 30, sgn(\mu) = +$.}}
\label{fig:pqcoup}
\end{figure}

\begin{figure}
\centerline{\includegraphics[angle=-0, width=.50\textwidth]{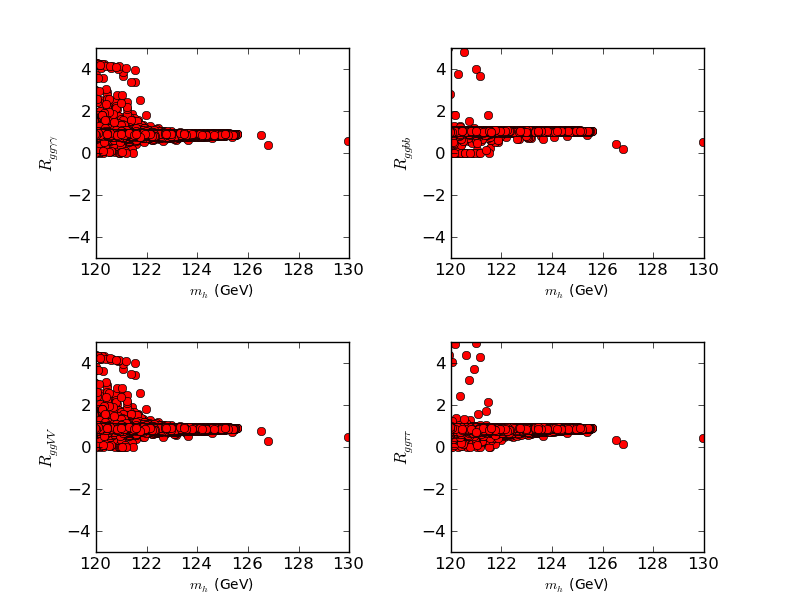}}
\caption{\color{black!100}{\sf Various Higgs-Boson observables, $R_{ggXY} = \frac{\sigma_{pp\to h} \times{\Gamma}^{h\to X Y}}{\Gamma_h}|_{\frac{SUSY}{SM}}$
 with respect to $m_h$ at the LHC.}}
\label{fig:pqratemh}
\end{figure}

\begin{figure}
\centerline{\includegraphics[angle=-0, width=.50\textwidth]{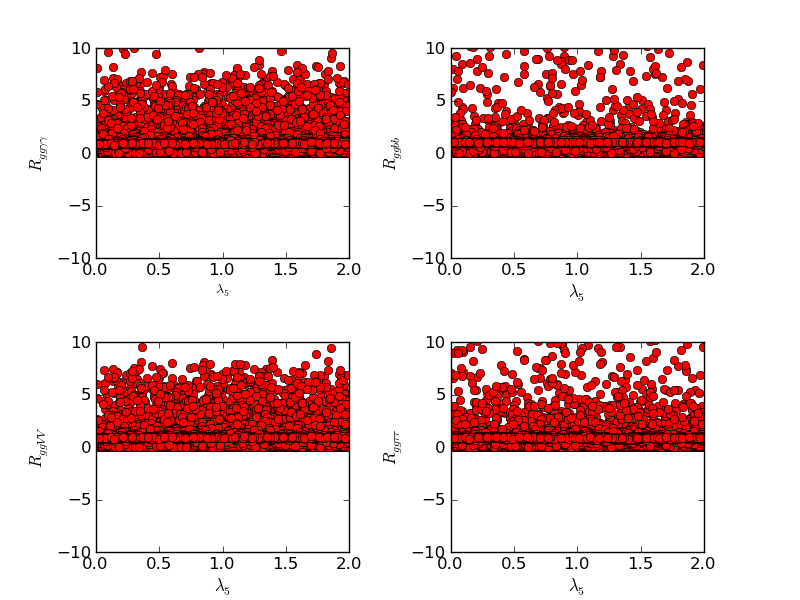}}
\caption{\color{black!100}{\sf Various Higgs-Boson observables, $R_{ggXY} = \frac{\sigma_{pp\to h} \times{\Gamma}^{h\to X Y}}{\Gamma_h}|_{\frac{SUSY}{SM}}$
 with respect to $\lambda_{5}$ at the LHC.}}
\label{fig:pqratela}
\end{figure}

\subsection{\label{pq:higgs} Implications for the Higgs Sector}

With a broken PQ symmetry the MSSM Higgs Lagrangian can be extended by the following term
\begin{eqnarray}
 -{\mathcal L} \supset \frac{\lambda_5}{2}(H_d^\dag H_u)^2 + h.c. ,
\label{eq.:hsuper}
\end{eqnarray}
where $\left| \lambda_5\right| \leq 2$.  As observed by the authors of  Ref. ~\cite{blum}, the above term modifies couplings of the Higgs-Boson to various SM particles by, 
\begin{subequations}
\begin{eqnarray}
\label{eq.:hcoupa}
 \delta g_{hb\bar{b}} &=& -  y_b \frac{r_t}{{\left(1 - m^2_h/M^2_{H_d}\right)}^2}\xi_{35} , \\
\label{eq.:hcoupb}
 \delta g_{ht\bar{t}} &=& y_t \left(\frac{M^4_{12}}{2 M^4_{H_d}}\right)\left (1 - \xi_{35}\right)\xi_{35} , \\
\label{eq.:hcoupc}
 \delta g_{hVV} &=& - \frac{2 M^2_V}{v}\left(\frac{M^4_{12}}{2 M^4_{H_d}}\right) \xi^2_{35} .
\end{eqnarray}
\end{subequations}
Here 
\begin{subequations}
\begin{eqnarray}
 \xi_{35} = \frac{\lambda_{35} v^2_u}{m^2_h - M^2_{H_d}} , \\
 \lambda_{35} = - \frac{g^2 + g^{\prime2}}{4} + \lambda_5 , \\
 r_t =\frac{g_{ht\bar{t}}}{y_t} .
\end{eqnarray}
\end{subequations}
From Eqs.(\ref{eq.:hcoupa}-\ref{eq.:hcoupc}) it is clear that the coupling of the SM-like Higgs to fermions with $T_3 = +1/2$ receives a positive correction due to PQ violation, while for $T_3 = -1/2$ fermions and the weak-bosons this contribution is negative.

\subsection{\label{pq:cosmo} Cosmological Implications} 

Due to the bounds on the PQ breaking scale in Eq.(\ref{Eq:LambdaPQBounds}), the axion mass is always restricted in the range between $\sim 1 \times 10^{-5}$ eV and $\sim 6 \times 10^{-3}$ eV.  The saxion is ultra-heavy and hence is less interesting for our purposes.  The axino mass can take values between 2 GeV and 1 TeV for the case $k=2$, and $\sim 1$ eV - $\sim 7$ keV for $k=3$.  Thus the axion is always the lightest of the three and hence serves as a good hot dark matter candidate.  The axino, covering a wide range of mass between a few eV to about a TeV, can be lighter, degenerate or heavier than the lightest neutralino.  This makes the PQ violating SuGra scenario very intriguing since the both the axino and the neutralino can contribute to the cold matter abundance.
For $\kappa = 2$, for example, three interesting PQSuGra scenarios are possible depending on the lightest neutralino mass, $m_{\widetilde{\chi}_1}$ and the axino mass, $m_{\tilde{a}}$. These are
\begin{itemize}
\item{\bf PQ-1} (neutralino LSP):  $m_{\widetilde{\chi}^0_1} < m_{\tilde{a}}$,
\item{\bf PQ-2} (axino-neutarlino co-LSPs): $m_{\tilde{a}} \simeq m_{\widetilde{\chi}^0_1}$,
\item{\bf PQ-3} (axino LSP): $m_{\tilde{a}} < m_{\widetilde{\chi}^0_1}$.
\end{itemize}
In the first (third) case the lightest neutralino $\tilde{\chi}^0_1$ (axino $\tilde{a}$) is the dark matter candidate.  In the second case they are both CDM candidates.

For $\kappa > 2$ the axino is always the lightest superpartner.  However, it was found that the only case that is cosmologically viable for $\kappa > 2$ is the one with $\kappa = 3$.  Scenarios with $\kappa > 3$ are unable to generate sufficient dark matter relic density and hence are less interesting.  Therefore in the current work we will only analyse the PQSuGra scenarios with $\kappa = 2$ (denoted by {\bf PQ-1, PQ-2, PQ-3}) and  $\kappa = 3$ ({\bf PQ-$3^\prime$}).

Following Ref. \cite{pqrelic1, pqrelic2}, we calculate the relic densities for the axions and axinos by the following formulae
\begin{subequations}
\begin{eqnarray}
 \Omega_a h^2 \simeq {1\over 4}\left(\frac{6\times 10^{-6}\ {\rm eV}}{m_a}\right)^{7/6} , \\
 \Omega_{\tilde a}^{NTP}h^2 = \frac{m_{\tilde a}}{m_{\widetilde{\chi}_1^0}}\Omega_{\widetilde{\chi}_1^0}h^2 , \\
\begin{split}
 \Omega_{\tilde a}^{TP}h^2 \simeq 5.5~g_s^6\ln\left(\frac{1.211}{g_s}\right) \times \\
 \left(\frac{10^{11}\ {\rm GeV}}{{\Lambda_{PQ}}/N}\right)^2 \left(\frac{m_{\tilde a}}{0.1\ {\rm GeV}}\right)\left(\frac{T_R}{10^4\ {\rm GeV}}\right) .
\end{split}
\end{eqnarray}
\end{subequations}
Here, $\Omega_a h^2$ is the axion relic density, 
$\Omega_{\tilde a}^{NTP}h^2$ is the relic abundance of non-thermally produced axinos from neutralino decay and $\Omega_{\tilde a}^{TP}h^2$ is the relic abundance of thermal produced axinos.

Axino dark matter lends the PQSuGra model considerably more viability compared to the minimal SuGra model.  The properties of axino dark matter, such as its abundance and couplings to standard matter, are governed by its mass and $\Lambda_{PQ}$ which are independent from the mSuGra parameters.  Thus, one expects more  flexibility from a model with more parameters.  Obtaining a Higgs mass of about 126 GeV, a neutralino abundance of 0.22 $\rho_C$ and low fine tuning is impossible in mSuGra.  The PQSuGra model not only accommodates these requirements much easier but its Bayesian evidence suggests that, despite of its extra parameters, it is more viable.


\section{\label{pq:stat} Numerical Analysis of the PQSuGra model}

In the rest of this paper we consider a mSuGra model extended with the PQ violating Higgs coupling Eq.(\ref{eq.:hsuper}).  The former is parametrized by the usual four parameters and a sign, while the latter adds two more to our full set of parameters:
\begin{eqnarray}
 \mathcal{P} = \left\{m_0, m_{1/2}, A_0, \tan\beta,  sgn(\mu), \lambda_5, \Lambda_{PQ}\right\} .
\label{Eq:paramscan}
\end{eqnarray}
We use the latest version of {\tt SUSY-HIT}~\cite{susyhit} to calculate sparticle masses and decay rates, and {\tt MicrOmegas 2.4.5}~\cite{micromegas} to calculate the relic density of lightest neutralinos.  To incorporate the effect of PQ violation we modified the relevant parts of SUSY-HIT.

At the leading one-loop level, the mass of the lighter CP-even Higgs-boson can be well approximated by the following equation~\cite{Haber:1996fp}:
\begin{widetext}
\begin{eqnarray}
m_{h}^2 &\simeq& \frac{1}{2}\left[M_A^2 + M_Z^2 - \sqrt{\left(M_A^2 + M_Z^2\right)^2 - 4 \cos^22\beta M_A^2M_Z^2}\right]  + \epsilon, {~\rm with}, \nonumber\\
\epsilon &=& \frac{3 m_t^4}{2\pi^2 v^2 \sin^2\beta}\left[\ln \frac{M_S^2}{m_t^2} + \frac{X_t^2}{2 M_S^2}\left(1 - \frac{X_t^2}{6 M_S^2}\right)\right] .
\end{eqnarray}
\end{widetext}
Here, $M_S = \frac{1}{ 2}(m_{{\tilde t}_1} + m_{{\tilde t}_2})$, $m_A$ is mass of the CP-odd Higgs boson, $m_{{\tilde t}_{1,2}}$ are the masses of the superpartners of top quark, and $X_t = A_t - \mu \cot\beta$ is the parameter that governs mixing in the stop-sector. Clearly $m_h$ increases with the paramters $m_A$, $m_{{\tilde t}_{1,2}}$, $\tan\beta$, $X_t$ ( or effectively $A_t$). This, in our case, translates into having larger values of $m_0$ and ${\cal A}_0$ along with a sufficiently large value of the parameter $\tan\beta$.

Due to the extra PQ violating term the Higgs couplings to various SM particles receive significant contribution compared to the PQ conserving mSuGra case.  This, in turn, modifies the Higgs branching ratios.  To show this, in Figure \ref{fig:pqcoup} we plot the deviation of the various Higgs branching ratios and the total decay width from the mSuGra ones.  We show this deviation as the function of $\lambda_{5}$ for fixed values of $\{m_0, m_{1/2}, A_0, \tan\beta,  sgn(\mu)\} = \{1000{~\rm GeV}, 500 {~\rm GeV}, 0 {~\rm GeV}, 30, +1 \}$. 

As expected the deviation rises with increasing $\lambda_{5}$ and becomes maximal at $\lambda_{5} = 2$.  Due to this the total Higgs decay width, for example, becomes about $10 (18) \%$ smaller for $\lambda_{5} = 1 (2)$ compared to the mSuGra case (which is equivalent to $\lambda_{5} = 0$).  This happens because for $\lambda_{5} > 0$ the $h-b-{\bar b}$ coupling becomes smaller.  After folding in changes of other Higgs couplings the $h \to \gamma \gamma$  branching ratio becomes larger by about $6 (18) \%$ for $\lambda_{5} = 1 (2)$compared to mSuGra.  There are similar changes to the $h\to V V^\star$, where $V = W^\pm, Z$ branching ratios.  The $h \to g g$ branching ratio increases by about $8 (19) \%$ for  $\lambda_{5} = 1 (2)$.  This translate into about $(100 - (100 + 8)\times (100 - 10))= 2.8 \%$, and, $(100 - (100 + 18)\times (100 - 18))= 3.2 \%$ decrease in the $h \to g g$ partial decay width for $\lambda_{5} = 1$, and, $2$ respectively.  Since Higgs is dominantly produced through the gluon fusion, the ratio 
\begin{eqnarray}
 \frac{\sigma^{PQSuGra}(p p \to h)} {\sigma^{mSuGra}(p p \to h)} \simeq \frac{\Gamma^{PQSuGra}(h \to g  g)} {\Gamma^{mSuGra}(h \to g g)} ,
\end{eqnarray}
also decreases accordingly for Higgs production at the LHC.  It is also clear from the plot that, for the selected parameter point, an overall increase of the diphoton event rates by about $1.03-1.14$ is predicted by PQSuGra over mSuGra.

As we saw, in the PQSuGra scenario it is possible to enhance the diphoton production rate and dark matter is less constraining than in mSuGra.  In the light of the LHC excess and the strict WMAP dark matter abundance constraint we aim to quantitatively compare the feasibility of PQSuGra and mSuGra.  In what follows we calculate Bayesian evidences for both models and evaluate their odds compared to each other.  Beyond the LHC Higgs search and WMAP we also use data from LEP, the Tevatron, and various low-energy experiments. 

The calculation of evidences involves a full scan over the parameter spaces of both models.  Motivated by naturalness, this scan is done over the following parameter ranges:
\begin{itemize}
\item $m_0 \in \left[10, 2000\right]$ GeV 
\item $m_{1/2} \in \left[10, 2000\right]$ GeV
\item  $A_0 \in \left[-3000, 4000\right]$ GeV
\item $\tan\beta \in \left[0, 62\right]$
\item  $\lambda_{5} \in \left[0, 2\right]$
\item  $\Lambda_{PQ} \in \left[1\times10^9, 5\times 10^{11}\right]$ GeV.
\end{itemize}

\subsection{\label{pq:bstat} Bayes Factor Calculation}


To assess the relative viability of PQSuGra compared to mSuGra we calculate their relative Bayes factor which is the ratio of their evidences.  First, for each observables ${\cal O}_i$ we calculate $\chi_i^2$ using the predicted and experimentally measured central values, ${\cal O}^{th.}_i$ and ${\cal O}^{exp.}_i$, and the associated uncertainties $\sigma_i^2 = (\sigma_i^{th.})^2 + (\sigma_i^{exp.})^2$:
\begin{eqnarray}
 \chi_i^2 = 
 \frac{\left({\cal O}^{th.}_i - {\cal O}^{exp.}_i\right)^2}{\sigma_i^2} .
\end{eqnarray}
The likelihood function can then be constructed as 
\begin{eqnarray}
{\cal L}_i = \frac{1}{{\sqrt{2 \pi} \sigma_i}} e^{-\chi^2_i/2} .
\end{eqnarray}
Since theoretical predictions ${\cal O}^{th.}_i$ depend on the theoretical parameters listed in Eq.(\ref{Eq:LambdaPQBounds}), the likelihood ${\cal L}_i$ carries the same dependence.

In terms of the likelihood the posterior probability distribution for parameter $p_j$ is then given by
\begin{eqnarray}
 P(x_j) = \frac{\displaystyle\int_{\{x \not = x_j\}}^{}{\cal P}(\{x\}){\cal L} \displaystyle\prod_{x \not= x_j}^{}dx}{\int_{\{x\}}^{}{\cal P}(\{x\}){\cal L} \displaystyle\prod_{x}^{}dx} ,
\end{eqnarray}
where ${\cal L} = \displaystyle\prod_{i}^{} {\cal L}_i$.  An $a$ $priori$ probability distribution ${\cal P}(\{x\})$ is assumed for each parameters.  While the PQSuGra parameters are listed in Eq.(\ref{Eq:LambdaPQBounds}), the mSuGra parameter space is smaller:
\begin{eqnarray}
 \mathcal{P} = \left\{ m_0, m_{1/2}, A_0, \tan\beta, sgn(\mu) \right\} .
\label{Eq:mSuGraPara}
\end{eqnarray}

The evidence $\cal E$ for each model is calculated by integrating the posterior density over all input parameters.  To give a specific meaning to the evidence, the Bayes factor is constructed:
\begin{eqnarray}
 Q_{Bayes} = \log_{10}\left(\frac{{\cal E}_{PQSuGra}}{{\cal E}_{mSuGra}}\right)
\end{eqnarray}
This evidence ratio quantifies the odds of PQSuGra against mSuGra.  Odds are interpreted in terms of the Jeffreys scale as shown in Table~\ref{t:jeff}
\begin{table}
\begin{tabular}{c c}
\hline
$Q_{Bayes}$ & Evidence against the  base model�� \\ \hline
$0 - 0.5$���� & Not worth mentioning \\ 
$0.5 - 1$���� & Substantial����������������������� \\ 
$1 - 2$�� ���� & Strong���������������������������� \\ 
�$�������>2$����& Decisive�������������������������� \\
\hline
\end{tabular}
\caption{Interpretation of Bayes factors on Jeffery's Scale.}
\label{t:jeff}
\end{table}

\subsection{Experimental Constraints}

Our likelihood function includes the following experimental and observational bounds on various observables and the (s)particle  masses.

\subsubsection{Precision observables (POs)}

\begin{itemize}
\item $\delta\rho = 0.0008 \pm 0.0017$~\cite{rho},
\item $a^{SUSY}_\mu = (3.353\pm 8.24) \times 10^{-9}$~\cite{amu},
\item $BR(b\to s \gamma) = (3.55\pm 0.26 \pm 5\% (th.)) \times 10^{-4}$~\cite{bsg}.
\end{itemize}

\subsubsection{Bounds on sparticle and Higgs boson mass from LEP-2/Tevatron}

\begin{itemize}
\item $m_{{\widetilde\chi}^0_{_1}} > M_Z/2$ GeV~\cite{rho, lepbounds},
\item $m_{{\widetilde\chi}^\pm_{_1}} > 103.5$ GeV~\cite{lepbounds},
\item $m_{\tilde l} > 99$ GeV~\cite{rho},
\item $m_h > 114.4$ GeV~\cite{lephiggs}.
\end{itemize}

\subsubsection{LHC data on $\sqrt{s} = 8$ TeV Higgs searches}

\subsubsection{LHC Higgs search}

We use the following LHC Higgs search observables
\begin{eqnarray}
 {\cal R}_{gg\gamma\gamma} &=& \left.\frac{\sigma_{pp\to h} \times{\Gamma}^{h\to \gamma\gamma}}{\Gamma_h}\right|_{\frac{SUSY}{SM}} , \nonumber \\
 {\cal R}_{gg2l2\nu} &=& \left.\frac{\sigma_{pp\to h} \times{\Gamma}^{h\to W^\pm W^\mp}}{\Gamma_h}\right|_{\frac{SUSY}{SM}} , \nonumber \\
 {\cal R}_{gg4l} &=& \left.\frac{\sigma_{pp\to h} \times{\Gamma}^{h\to ZZ}}{\Gamma_h}\right|_{\frac{SUSY}{SM}} .
\end{eqnarray}
Here ${\cal R}_{gg\gamma\gamma}, {\cal R}_{gg2l2\nu}, {~\rm and}, {\cal R}_{gg4l}$ are ratios of diphoton, $2l 2\nu$ (where $l = e^\pm, \mu^\pm$), and $4l$ production rates in PQSuGra relative to the Standard Model.  We did not include ${\cal R}_{gg\tau^\pm\tau^\mp}$ because there is insufficient data corresponding to this observable. 

The following are the experimental values used, given in the Ref.~\cite{cms},
\begin{itemize}
\item $m_h = 126 \pm 0.5657 (exp.) \pm 2 (th.)$ GeV
\item $R_{gg\gamma\gamma} = 1.8\pm 0.5$ 
\item $R_{gg2l2\nu} = 1.4\pm 0.5$ 
\item $R_{gg4l} = 1.4\pm 0.6$
\end{itemize}

\subsubsection{Dark matter abundance}

\begin{itemize}
\item $\Omega_{DM} = 0.1123 \pm 0.0035 \pm  10\% (th.)$~\cite{wmap}
\end{itemize}

In order to estimate various LHC observables we calculate the SM and SUSY Higgs decay width and branching ratios at NNLO + NNLL, wherever available.  Numerical values for other observables (except for relic-density, $\Omega_{DM}$) were also calculated using {\tt SUSY-HIT}.  For relic density predictions we use {\tt MicrOmegas}.  

To safely avoid a charged LSP we assume that masses for all the sleptons and squarks are at least 10 GeV above the axino and/or neutralino LSP masses. 

Before closing this section we would like to add that for the numerical implementation of various PQSuGra scenarios, we assume the mass difference between the co-LSPs is within 10\%.

\section{\label{pq:res} Results}

The $\sqrt{s} = 8$ TeV LHC Higgs decay observables show a significant deviation from the SM prediction particularly for the diphoton, di-lepton and the four lepton cases where the observed central value for the former turns out to be about 1.5-1.8 times larger than the SM.  We have shown that PQSuGra can change these observables significantly in the right direction compared to the mSuGra.  To gain detailed insight in Figs.~\ref{fig:pqratemh} we plot $R_{gg\gamma\gamma}$, $R_{ggbb}$, $R_{ggVV}$, and $R_{ggVV}$ with respect to the Higgs mass for PQSuGra.  These plots are the result of a scan over the PQSuGra parameter ranges given above.  The plots tell that for most of the cases the diphoton event rate drops considerably with increasing Higgs mass.  Yet there still seems to be a narrow region where the Higgs mass and $R_{gg\gamma\gamma}$ agrees with the data within $2\sigma$.  (We assume about 1 GeV theoretical uncertainty in the Higgs-Boson mass calculation.)  The other two observables, $R_{gg2l2\nu}$ and $R_{gg4l}$, are also in good agreement with the LHC measurements.  To show how sensitive each of these observables are for the PQ violating effects, in Figs.~\ref{fig:pqratela} we plot them with respect to $\lambda_5$. This plot tells that it is possible to satisfy the LHC data for a wide range  of the $\lambda_5$. 

\begin{figure}
\centerline{\includegraphics[angle=-90, width=0.45\textwidth]{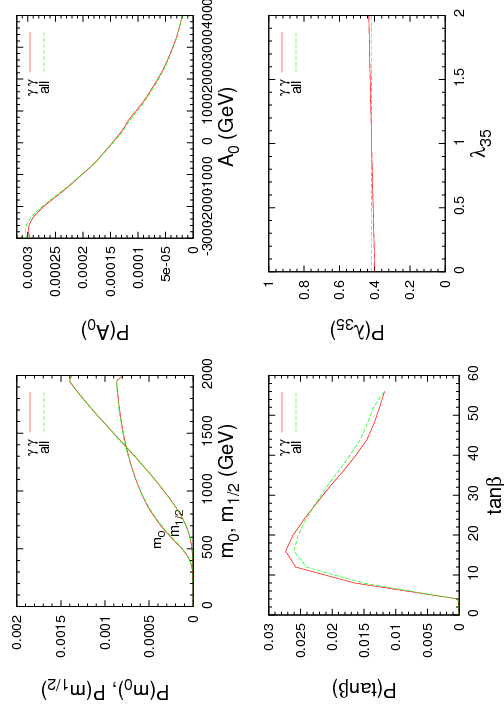}}
\caption{\sf Posterior probability distributions using the flat prior.}
\label{fig:post_grid}
\end{figure}

\begin{figure}
\centerline{\includegraphics[angle=-90, width=0.45\textwidth]{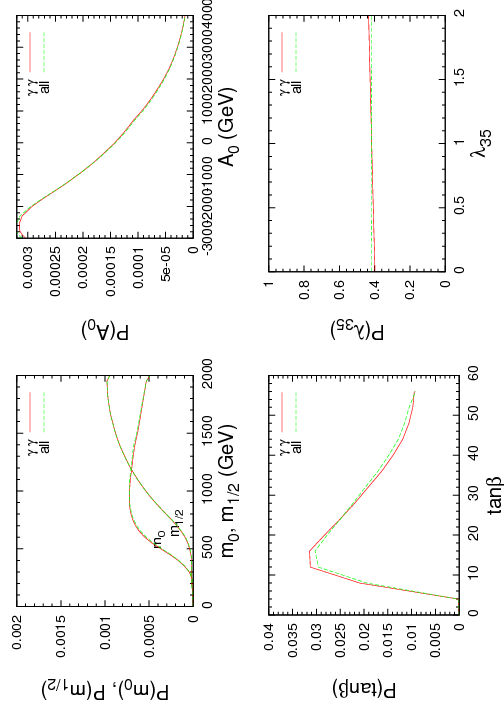}}
\caption{\sf Posterior probability distributions using the natural prior.}
\label{fig:post_grid_log}
\end{figure}

We calculate posterior probabilities for the SuGra and PQSuGra models for two popular choices of priors: (a) the flat (uniform) prior which is constant in a finite parameter region; and (b) the log prior $\propto (m_0 m_{1/2})^{-1}$.  These posterior probability distributions are then marginalized to the various parameters and plotted in Figs.~\ref{fig:post_grid} and ~\ref{fig:post_grid_log}.  As these figures show a large value of $m_{0} > 1.5$ and  $m_{1/2} > 0.5$ TeV, $A_0 < -3$ TeV and $\tan\beta \sim 20$ is preferred by the experimental data.  

The integral of the posterior probability distributions allows us to calculate the Bayes factors as presented in Table~\ref{tab:qbayes}.  This table along with Table~\ref{t:jeff} shows that the PQSuGra scenario is somewhat better then the mSuGra when only Higgs and POs are imposed.  However, scenarios where the axino is dark matter are strongly preferred over mSuGra.  This is because in the latter it is hard to satisfy the LHC Higgs mass constraint, WMAP and $g-2$ simoultaneously.  In contrast, the $PQ-3$ (and $PQ-3^\prime$) models satisfy WMAP in a wider range of the PQSuGra parameter space according to Figs.~\ref{fig:pqratela}.


\section{\label{pq:concl} Conclusions}

We have preformed a Bayesian analysis of the minimal and Peccei-Quinn violating supergravity models.  We compared the viability of the two models in light of the LHC Higgs searches at $\sqrt{s} = 8$ TeV, the WMAP data on the relic density of dark matter of the Universe, along with data from various other experiments.  Our study reveals that PQSuGra scenarios with an axino LSP are clearly preferred by the collider and astrophysical data.  

\begin{figure}
\centerline{\includegraphics[angle=-00, width=.40\textwidth]{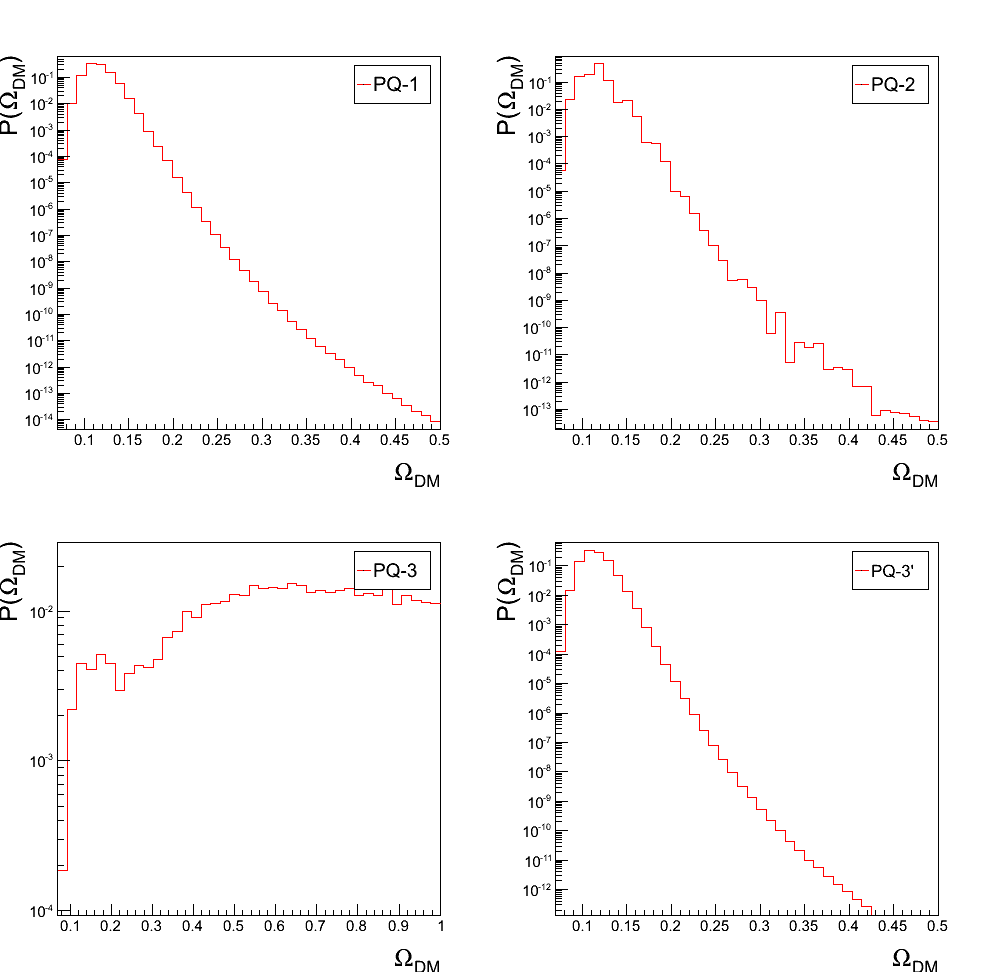}}
\caption{\sf Posterior probability distributions for the relic density ($\Omega_{DM}$) for various PQSuGra scenarios.}
\label{fig:post_omega}
\end{figure}
\begin{table}
\resizebox{9.0 cm}{!} {
\begin{tabular}{ | l | l | l |l| l|}\hline
\emph{Observables}& \multicolumn{4}{|c|}{\emph{$Q_{Bayes}$}} \\
\cline{2-5}
 & \emph{PQ-1} & \emph{PQ-2} & \emph{PQ-3} & \emph{PQ-3$^\prime$} \\\hline\hline
Higgs Searches at the LHC & 0.244 & 0.305 & 0.663 & 0.506 \\\hline
+ POs \& LEP data        & 0.238 & 0.322 & 0.724 & 0.626 \\\hline
+ WMAP data               & 0.181 & 0.231 & 1.693 & 2.466 \\\hline
\end{tabular}
}
\caption{\sf Bayes factors for various PQSuGra scenarios for $m_0 \in \left[10, 2000\right]$, $m_{1/2}
\in \left[10, 2000\right]$, $A_0 \in \left[-3000, 4000\right]$ (all in GeV units), $\tan\beta \in \left[0, 62\right]$, $\lambda_{5} \in \left[0, 2\right]$, $\lambda_{PQ} \in \left[1 \times 10^{9},
5\times 10^{11}\right]$ (in GeV).}
\label{tab:qbayes}
\end{table}


\begin{acknowledgments}

We thank Ben Farmer for useful discussions on Bayesian statistics and Philip Chen for his assistance with the cluster computing.  This work was supported in part by the {\em ARC Centre of Excellence for Particle Physics at the Tera-scale}.  The use of Monash University Sun Grid, a high-performance computing facility, is gratefully acknowledged.

\end{acknowledgments}

\end{document}